# Impact of Contact Gating on Scaling of Monolayer 2D Transistors Using a Symmetric Dual-Gate Structure

*Victoria M. Ravel[1], Sarah R. Evans[1], Samantha K. Holmes[1], James L. Doherty[1], Md Sazzadur Rahman[1], Tania Roy[1], and Aaron D. Franklin[1,2]\**

[1]Department of Electrical & Computer Engineering, Duke University, Durham, NC, USA

[2]Department of Chemistry, Duke University, Durham, NC, USA

\*Address correspondence to aaron.franklin@duke.edu

## Abstract

The performance and scalability of two-dimensional (2D) field-effect transistors (FETs) are strongly influenced by geometry-defined electrostatics. In most 2D FET studies, the gate overlaps with the source/drain electrodes, allowing the gate potential to modulate the 2D semiconductor underneath the electrodes and ultimately effect carrier transport at the metal-semiconductor interface—a phenomenon known as contact gating. Here, a symmetric dual-gate structure with independently addressable back and top gates is employed to elucidate the impact of contact gating on a monolayer MoS$_2$ channel. Unlike previous studies of contact gating, this symmetric structure enables quantification of the phenomena through a contact gating factor ($\beta_{CG}$), revealing a ~2× enhancement in on-state performance in long-channel devices. At scaled dimensions (50 nm channel and 30 nm contact length), the influence of contact gating becomes amplified, yielding a ~5× increase in on-state performance and a ~70% reduction in transfer length when contact gating is present. Since many reported record-performance 2D FETs employ back-gate geometries that inherently include contact gating, these results establish contact gating as a critical and previously underappreciated determinant of device performance in the 2D FET landscape.

**Introduction and Background**

The potential of atomically thin two-dimensional (2D) semiconducting transition metal dichalcogenides (TMDs) for next-generation, aggressively scaled field-effect transistors (FETs) is well recognized.[1–4] The International Roadmap for Devices and Systems (IRDS) identifies 2D TMDs as key enablers of future semiconductor technologies, particularly as channel materials for ultra-scaled, low-power transistors.[5] Among these materials, monolayer $MoS_2$ has been established as the prototypical n-type TMD and has been the focus of the majority of scaling studies to date.[6–15] Significant progress has been made toward overcoming the key challenges to realizing high-performance 2D FETs, particularly through demonstrations of devices with ultralow contact resistance ($R_C$), approaching the quantum limit.[16,17]

However, the interpretation of these noteworthy demonstrations is often complicated, as device architectures in which the contact resistance is a function of the applied gate voltage are commonly employed.[13,18–21] This dependence arises when the gated region of the device overlaps with the source/drain (S/D) contacts, allowing the electrostatic field from the gate to modulate the 2D semiconductor in these regions. As a result, the gate potential alters the local band alignment and carrier concentration beneath the contacts, influencing the contact resistance—an effect referred to as contact gating (CG).[22–25]

CG is prevalent in back-gated geometries (often with an entire substrate used as the gate electrode), where the metal contacts are deposited on top of the 2D semiconductor. As shown in **Fig. 1ai**, this configuration results in an overlap between the S/D contacts and the back gate, enabling the gate electric field to modulate the 2D semiconductor underneath the contact regions (i.e., contact gating). To the best of our knowledge, CG is also present in all reported 2D semiconductor gate-all-around (GAA) structures to date, where the semiconducting channel sits on top of the S/D electrodes in a bottom-contacts configuration, and the top gate (unlike the self-aligned back gate) overlaps with the S/D electrodes to modulate the semiconductor in the contact regions.[26–28] Importantly, a device with CG is impractical for high-performance transistor technology, as it introduces substantial parasitic capacitance between the gate and the S/D contacts, resulting in increased power consumption and degraded intrinsic delays.[29] These parasitic effects often go unnoticed in DC or low-frequency characterization, leading to the misleading reporting of important benchmarking parameters.[23–25]

Several studies have investigated the influence of contact gating on 2D FET performance, identifying a significant impact on $R_C$[22,29,30] and transfer length ($L_T$)[22], as well as complicating the extraction of field-effect mobility ($\mu_{FE}$).[24,31] While it is now evident that CG greatly impacts device performance, its role remains poorly understood, particularly regarding its impact on scaled monolayer $MoS_2$ FETs.

In this work, we use a symmetric dual-gate structure, with independently addressable top and back gates, to quantitatively evaluate the contact gating effect in monolayer $MoS_2$ FETs. The symmetry of the top and back gates allowed for the direct extraction of a contact gating factor for both the off- and on-state regimes. Furthermore, by applying this analysis to devices with scaled channel and contact lengths ($L_{ch}$ and $L_c$), we demonstrate that the impact of contact gating becomes increasingly pronounced at scaled dimensions. These findings increase awareness of contact gating effects in 2D FETs and encourage a transition away from conventional back gate architectures toward device designs that more accurately reflect the intrinsic performance of the materials and meet the requirements of future high-performance transistor technologies.

**Fabrication of the Symmetric Dual-Gate Structure**

The cross-section schematic of the symmetric dual-gate monolayer $MoS_2$ FETs is illustrated in **Fig. 1a**. Devices were fabricated on intrinsic silicon substrates with a 1 μm thermal $SiO_2$ layer. The high resistivity of the intrinsic silicon (>20 kΩ·cm) ensures electrical isolation between the local back gates and the substrate. Local back gates (Ni/Au) were first patterned into the substrate, after which a 25 nm $HfO_2$ bottom dielectric was grown via thermal atomic layer deposition (ALD) at 200°C. A thin (~1 nm) Al seed layer was then deposited on top of the dielectric to ensure a symmetrical interface with the seed layer used later for the top-dielectric nucleation. The Al layers were naturally oxidized in air for 24 hours to form thin $Al_2O_3$ layers prior to subsequent processing.

The monolayer $MoS_2$ used in this study was synthesized by metal–organic chemical vapor deposition (MOCVD) and subsequently transferred onto the prepared substrate using a polymethyl methacrylate (PMMA)-assisted water transfer method. Details of the growth and transfer procedures are provided in the Methods section, and materials characterization is discussed in **Supplementary Note 1** and **Fig. S1**.

After transfer, the S/D contacts (Ni/Au) were patterned and deposited via ultra-high vacuum electron-beam evaporation (~$10^{-8}$ Torr). The $MoS_2$ channel width was then defined by reactive ion etching using $SF_6$ plasma. To complete the top gate stack, an Al seed layer was again deposited and oxidized in air, followed by the deposition of the $HfO_2$ top dielectric and top gate using the same processes employed for the bottom layers. The complete fabrication process is described in detail in the Methods section and illustrated in **Supplementary Fig. S2**.

The resultant device structure (**Fig. 1a**) provides consistent control of the $MoS_2$ channel from both gates, with their influence on the contact regions differing considerably. When biased through the back gate, the electric field modulates the $MoS_2$ underneath the contacts (i.e., contact gating) (**Fig. 1ai**). When the device is controlled through the top gate, the electric field is confined to the channel region and is shielded from modulating the MoS$_2$ within the contact regions due to the metal S/D contacts (**Fig. 1aii**). Though fringing fields from the top gate will contribute to some CG effects[32] this effect is negligible compared to the full CG resulting from the back gate.

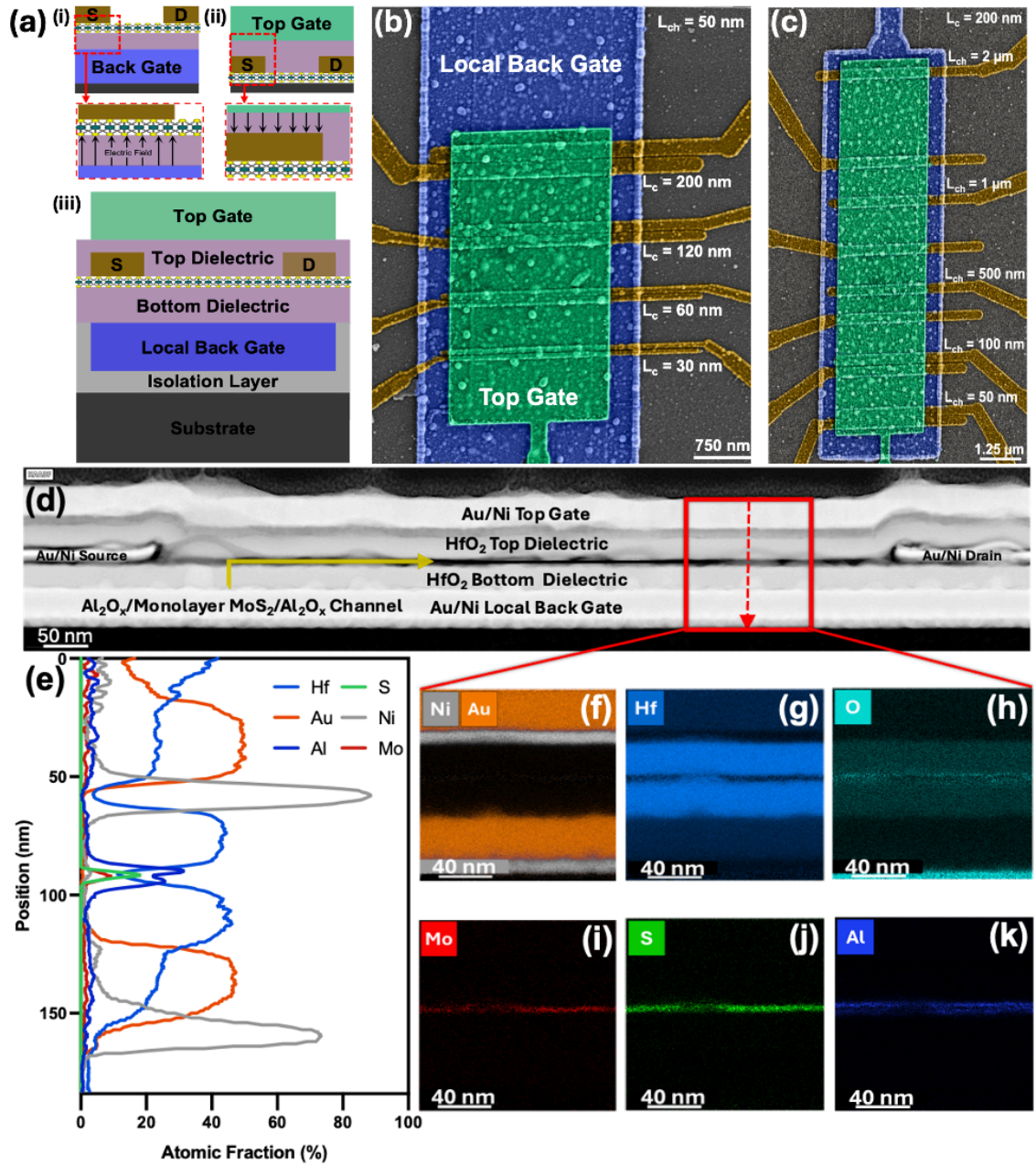

**Figure 1. Materials characterization of symmetric dual-gate structure.** (a) Cross-sectional schematics of the dual-gate structure, including: (i) Back-gated structure highlighting gate modulation of the MoS$_2$ underneath the contacts, (ii) top-gated structure without gate modulation underneath the contacts, and (iii) full device cross-section showing the symmetric dual-gate structure. Local back gate is blue, S/D contacts are brown, and top gate is green. False-colored SEM images of (b) contact length-scaled devices and (c) channel length-scaled symmetric dual-gate FETs with the same color scheme as (a). The patterned MoS$_2$ channel can be faintly seen, slightly smaller in width than the top gate. (d) Cross-sectional STEM image confirming the symmetrical dual-gate structure. EDS mapping shows symmetry between the back- and top gate stack with (e) atomic makeup of the device shown along the dotted red line and (f-k) elemental mapping confirming the locations of Ni, Au, Hf, Al, Mo, S, and O.

Sets of devices with varying channel lengths ($L_{ch}$, down to 50 nm) and contact lengths ($L_c$, down to 30 nm) were fabricated across the chip, with representative top-view scanning electron microscopy (SEM) images in **Figures 1b** and **1c**. Structural symmetry was confirmed using cross-sectional scanning transmission electron microscopy (STEM) combined with energy dispersive x-ray spectroscopy (EDS), with a representative STEM cross-section shown in **Fig. 1d** and the corresponding atomic makeup in **Fig. 1e** with the EDS elemental mappings in **Figures 1f-k**.

**Evaluation of the Contact Gating Effect**

For the electrical characterization of the devices in this work, one of the gates was swept to modulate the MoS$_2$ channel and control the device operation/performance (i.e., the control gate, $V_{GS}$), while the other was held at a fixed bias and served as the static gate ($V_{static}$). In this manuscript, all the blue data corresponds to sweeping the back gate as $V_{GS}$ (i.e., full CG effect) and all green data corresponds to sweeping the top gate as $V_{GS}$ (i.e., fixed/static CG dependent on $V_{static}$). Characteristics from a long-channel device ($L_{ch}$ = 1 µm, $L_c$ = 200 nm) provide a model example of the CG effect (**Fig. 2**). The transfer curves clearly demonstrate how contact gating enhances the on-state behavior: back gate sweeps (i.e., the blue curves) yield higher on-current ($I_{ON}$), larger transconductance ($g_m$), and a more expansive linear response across $V_{GS}$. The subthreshold and transfer characteristics of the top and back gates swept with $V_{static}$ = -5 V are compared in **Fig. 2a**. Both gates exhibit similar threshold voltages ($V_{th}$) around -3.8 V, but the back gate achieves an $I_{ON}$ of ~17 µA/µm, compared to ~7 µA/µm for the top gate — a ~60% decrease in $I_{ON}$ when switching from back gate to top gate control. The subthreshold and transfer characteristics of the back gate for $V_{static}$ values ranging from -5 V to 0 V in 0.5 V increments are shown in **Figures 2b** and **2c**, respectively, while the corresponding data for the top gate is in **Figures 2d-e**. The degradation in $g_m$ and linearity is particularly evident in the top gate-controlled devices when comparing **Figures 2c** and **2e**. Of note is that as the $V_{BG} = V_{static}$ becomes more positive in **Fig. 2e**, the transfer curve presents more similar to the curves in **Fig. 2c** owing to the increased influence of the $V_{BG}$ contact gating.

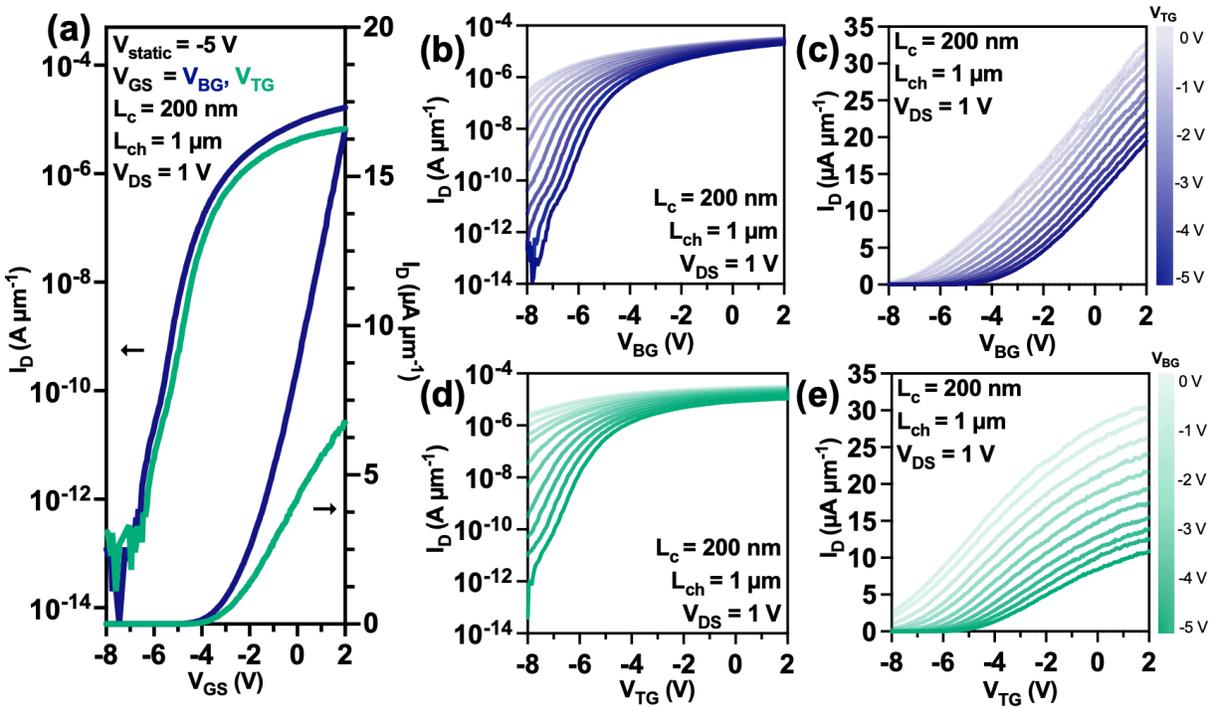

**Figure 2. Comparison of performance with and without contact gating** – blue curves are with contact gating (back gate control) and green curves without contact gating (top gate control) on the same monolayer MoS$_2$ channel with non-swept gate held at static voltage ($V_{static}$). (a) Subthreshold and transfer characteristics of $V_{BG}$ and $V_{TG}$ with static voltage of -5 V. (b) Subthreshold and (c) transfer characteristics of back gate swept with static top gate from -5 V to 0 V. (d) and (e) are the same as (b) and (c), respectively, with the gates swapped.

**Off-state and On-state Implications of Contact Gating**

If the dual gates influenced the channel identically, applying a static bias to one gate while sweeping the other would result solely in a $V_{th}$ shift in the transfer characteristics.[23] To evaluate the specific influence of CG in the off-state, the $V_{th}$ of the control gate was compared to the applied static gate bias (**Supplementary Fig. S3**).

In a perfectly symmetric device where both the top and back gates exhibited equal control over the channel, in the off-state, the relationship between $V_{th}$ and $V_{static}$ should yield a slope of $V_{th}:V_{static} = -1$, as any shift in the $MoS_2$ bands induced by the static gate would require an equal and opposite voltage from the control gate to reverse it, thus affecting the $V_{th}$.[23,30,33] However, in this device, while the gate stacks are symmetric in their modulation of the channel, the back gate uniquely modulates the contacts (**Fig. 3a**). Thus, the back gate will have a stronger influence on the threshold voltage than the top gate, and the $V_{th, BG}$ verse $V_{TG-static}$ relationship would deviate from -1 based on some contact gating factor ($\beta_{CG-off}$). For example, the extracted slope of $V_{th, BG}:V_{TG-static}$ for the long-channel device in **Fig. 2** is -0.64 (see **Fig. S3a**), corresponding to $\beta_{CG-off}$ = 1.56 (See **Fig. S3** for details). This indicates that the expected $\Delta V_{th, BG}$ due to a $\Delta V_{TG-static}$ is offset from the -1:1 relationship by a factor of $\frac{1}{1.56}$. Conversely, when the back gate is held static, the $V_{th, TG}:V_{BG-static} = -0.83$, as the $MoS_2$ beneath the contacts is no longer being modulated by a sweeping back gate, and thus the full CG effect is not present. However, a small CG effect remains because the fringing fields from the top gate are still present.

In the on-state, however, the impact of contact gating is much stronger. In the case of an ideal device with equal control from the top and back gates (i.e., no contact gating effect) or in the case of a device that had a static back gate (i.e., constant contact gating effect), increasing $V_{static}$ will simply change the threshold voltage. Therefore, the current at a fixed overdrive voltage, $V_{OV} = V_{GS} - V_{th}$, will not depend on $V_{static}$. However, evaluating $I_{ON}$ at a fixed overdrive for both the $V_{static, TG}$ and $V_{static, BG}$ cases, shown in **Fig. 3b**, highlights the boost in performance provided by CG. With $V_{BG}$ as the control gate, $I_{ON}$ (taken at a consistent $V_{OV}$) remains nearly invariant with changes in $V_{static,\,TG}$ due to full contact gating being present regardless of how $V_{static,\,TG}$ shifts the threshold voltage. In other words, the applied $V_{static,\,TG}$ has no measurable effect on the back gate-controlled on-current. In this regime, CG effectively thins the Schottky barrier for carrier injection, thereby reducing $R_C$[23], or more specifically, the junction resistance component of $R_C$.[34] Conversely, when the top gate is used as the control gate and the back gate is held static, the amount of modulation of the semiconductor underneath the contacts due to CG is also held static, and when $V_{static,\,BG} \gg V_{th,\,TG}$ (i.e., past the contact-limited region of the $I_D$-$V_{GS}$ relationship), $I_{ON}$ is effectively equivalent to what is achieved with the back gate control scenario, as seen in **Fig. 3b** when $V_{static} > 0$ V. However, as $V_{static,\,BG}$ moves closer to $V_{th,\,BG}$, $I_{ON}$ decreases approximately linearly. Assuming the on-state CG effect is minimized when $V_{static,\,BG} \approx V_{th,\,TG}$ (obtained through linear extrapolation from peak $g_m$ of the transfer curve), then the extracted on-state CG factor is $1.72 < \beta_{CG-on}(I_{ON}) < 2.22$. This indicates that contact gating enhances the $I_{ON}$ by roughly 2× in this $L_{ch}$ = 1 μm device. Using the same $V_{static\ BG/TG} \approx V_{th,\ TG/BG}$ assumption a $g_m$-based CG factor ($\beta_{CG-on}(g_m)$) was also extracted, yielding values in the range of 1.90 to 2.02, consistent with the $I_{ON}$ enhancement of ~2×(**Fig. 3c**). Additional dependencies of these $\beta_{CG-on}$ on drain-source voltage ($V_{DS}$) and $V_{static}$ are provided in **Fig. S4**.

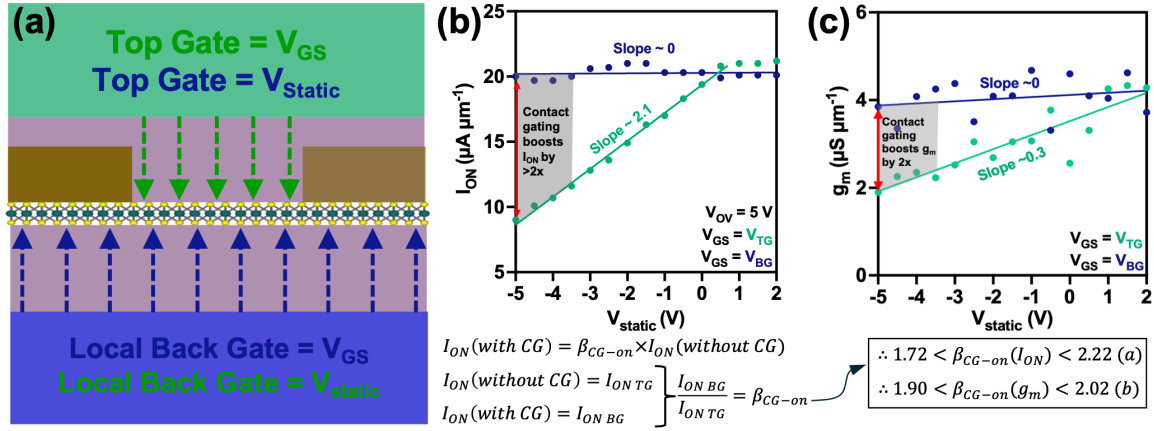

**Figure 3. On-state comparison of how $I_{ON}$ and $g_m$ are affected by contact gating** for the device in Fig. 2, including the extraction of a CG factor for the on-state, $\beta_{CG-on}$. (a) Cross-sectional graphic depicting the electric field difference between the bottom and top gates, highlighting the CG under the contacts for the local back gate. (b) $I_{ON}$ extracted at the same $V_{OV}$ for each static voltage. (c) Same as (a), but transconductance is extracted by averaging the slope over 11 points by taking 5 points above and below the steepest slope point on each transfer curve. Gray regions are where $V_{static} \approx V_{th,\,TG}$, representing an assumed minimization of CG in the green (top gate-controlled devices. $\beta_{CG-on}$ reveals that devices with contact gating receive a ~2x boost in on-state performance).

To evaluate $R_C$ dependence on sheet carrier density $n_s = \frac{C_{ox} \cdot V_{ov}}{q}$, transfer length method (TLM) plots[35] were constructed from sets of devices having different $L_{ch}$ (**Supplementary Note 2**). An SEM image of one such TLM structure can be seen in **Fig. 1c**. Although the relatively small number of $L_{ch}$ devices reduces the reliability of quantitative extractions from these TLM plots, the analysis clearly captures comparative trends between devices with and without CG. The extracted $2R_C$ values are relatively high (10's of kΩ-μm), but not inconsistent with many prior reports,[36–38] and still sufficient to reveal the direct impact of CG on $R_C$. At $V_{static}$ = -5 V, which is approximately equal to $V_{th}$, the contact resistance at high $n_s$ is ~2.5x larger in the absence of CG (**Fig. S5c**). When $V_{static}$ is shifted deeper into the subthreshold regime (-7 V), the $MoS_2$ in the contact regions of the top gate-controlled device becomes severely depleted,[22,39] leading to a $2R_C$ value nearly an order of magnitude higher than that of the contact-gated (back-gated) device (**Fig. S6c**). A complementary view of the CG effect is seen in **Fig. S7** and **Supplementary Note 3** with the $μ_{FE}$ versus $L_{ch}$ plotted at different $V_{static}$ values, where the influence of contact gating through variations in $R_C$ becomes evident. This result reflects the sensitivity of mobility extraction itself, which is frequently over- or under-estimated.[24,31] Moreover, in sub-300 nm $L_{ch}$ devices, the concept of mobility loses its physical significance as device performance is contact-limited.[40]

**Scaling and the Contact Gating Effect**

As the channel and contact lengths are scaled down, the relative contributions of channel and contact regions in determining performance of the 2D FET change accordingly. The transfer characteristics of various FETs as their $L_{ch}$ is scaled down from 2 μm to 100 nm are shown in **Fig. 4a-b**. The general increase in $I_{ON}$ observed in the characteristics can be attributed to the reduction in channel resistance ($R_{CH}$). However, this scaling also enhances the influence of CG as the overall device performance becomes contact dominated.

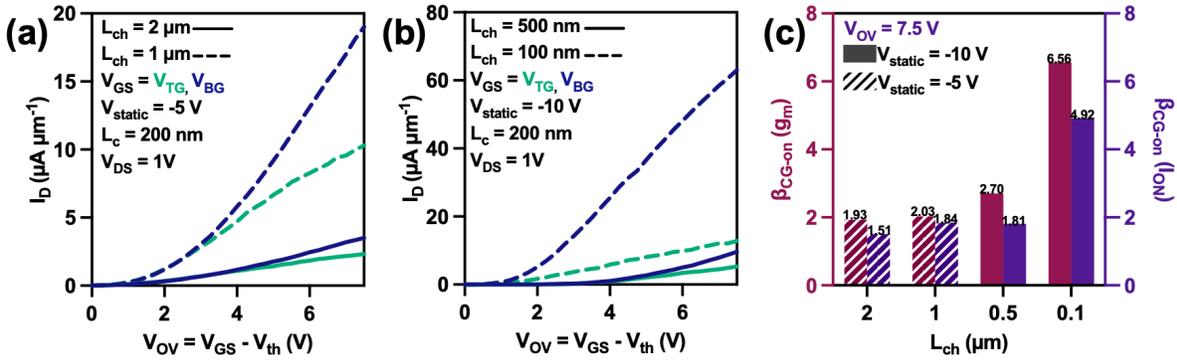

**Figure 4. Impact of channel length scaling on contact gating effect in the on-state.** $I_D$-$V_{OV}$ transfer characteristics of (a) 2 μm and 1 μm channel lengths at $V_{static}$ = -5 V and (b) 500 nm and 100 nm channel lengths at $V_{static}$ = -10 V (as with previous figures, blue data denotes back gate control and thus full CG, while green data denotes top gate control). Note, $V_{static}$ was chosen at approximately the same offset from $V_{th}$ for the respective devices. (c) $β_{CG-on}$ at various $L_{ch}$ values, showing an amplified impact on small-channel devices owing to more contact-limited performance.

When $L_{ch}$ is scaled to 500 nm, the on-state CG factors, $β_{CG-on}$ ($I_{ON}$) and $β_{CG-on}$ ($g_m$), remain relatively steady at ~1.8 and ~2.7, respectively (**Fig. 4c**). Further scaling $L_{ch}$ to 100 nm results in a substantial increase, with $β_{CG-on}$ ($I_{ON}$) and $β_{CG-on}$ ($g_m$) reaching ~4.9 and ~6.6, respectively (**Fig. 4c**). The CG effect is amplified because, as $L_{CH}$ scales down, so too does $R_{CH}$ until it becomes negligible compared to $2R_C$, shifting the device into a contact-dominated performance regime where the gate modulation of the contacts is more significant to the overall performance.

To further evaluate this regime, devices with $L_{ch}$ = 50 nm—where $R_{CH}$ is negligible[41]— were used to study $L_c$ scaling and its impact on the CG effect (**Fig. 5**). The transfer characteristics of these devices for $L_c$ values down to 60 nm under both back- and top gate control with $V_{static}$ = -10 V. Consistent with the $L_{ch}$ scaling results, the initial $β_{CG-on}$ values at $L_c$ = 200 nm were ~4-5 for both $I_{ON}$ and $g_m$, as the smallest $L_{ch}$ device in the $L_{ch}$ scaling structures is the same as the largest $L_c$ device in the $L_c$ scaling structure in terms of channel and contact lengths (See **Fig. 1b,c** for reference). As $L_c$ decreased to 60 nm, $\beta_{CG\text{-}on}$ increased to ~6.3 and ~7.3 for $I_{ON}$ and $g_m$, respectively (**Fig. 5b**). A complementary view of this trend is shown in **Fig. 5c** for a device with $L_{ch}$ = 50 nm and $L_c$ = 30 nm, characterized with and without CG. This device yields $4 < \beta_{CG\text{-}on}$ ($I_{ON}$) $< 4.87$, directly comparable to the large device, as it was measured the same $V_{OV}$ and $V_{static}$ values, which exhibited $\beta_{CG\text{-}on}$ ~ 2 (**Fig. 3a**). The corresponding transfer characteristics for top- and back gate control of this device are provided in **Fig. S8**.

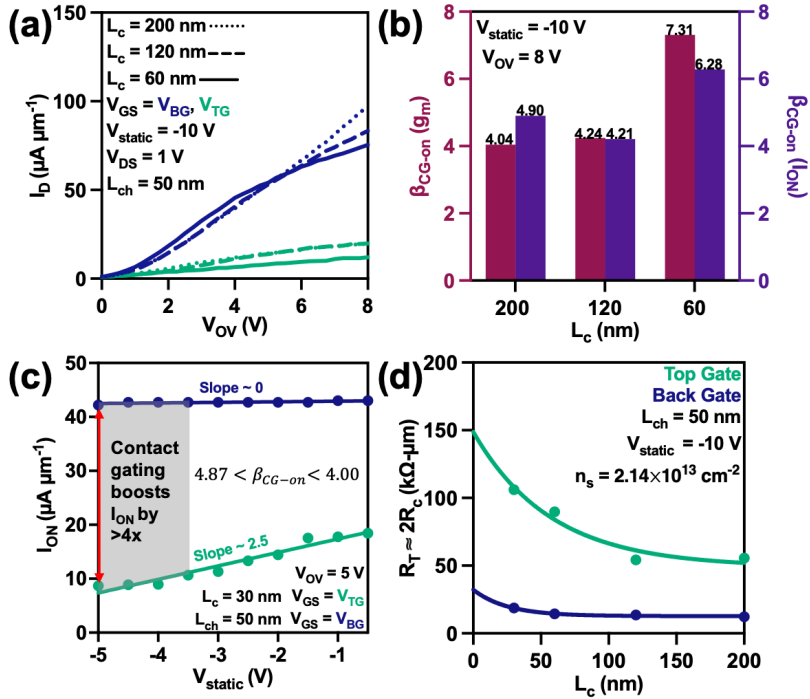

**Figure 5. Impact of contact length scaling on the contact gating effect in the on-state.** (a) $I_D$-$V_{OV}$ transfer characteristics of 200 nm, 120 nm, and 60 nm contact lengths at $V_{static}$ = -10 V, comparing top (green) versus back gate control (blue). (b) $\beta_{CG-on}$ ($I_{ON}$) and $\beta_{CG-on}$ ($g_m$) at the same $L_c$ values as (a), showing a significant increase at scaled $L_c$ indicative of the influence of CG on contact scaling. (c) $I_{ON}$ extracted at the same $V_{OV}$ for each $V_{static}$. Gray region is $V_{static} \approx V_{th, TG}$, representing an assumed minimization of CG in the green (top gate controlled) devices. Contact gating yielded a >4x boost in on-state performance; this is directly comparable (at the same $V_{OV}$ and $V_{static}$ values) to Fig. 3b, demonstrating an increase in $\beta_{CG-on}$ by a factor of ~2 when $L_c$ and $L_{ch}$ are both scaled. (d) $R_T$ vs. $L_c$ at $n_s$ = 2.14×10$^{13}$ cm$^{-2}$, revealing a distinct difference in $L_c$ scaling behavior with contact gating present (blue), yielding a ~70 % smaller transfer length.

In devices with scaled contacts, the length over which most of the current is injected into the semiconductor (the transfer length, $L_T$) can exceed the physical contact length ($L_T > L_C$). When this occurs, the $R_C$ increases exponentially with reduced $L_C$ as current injection becomes bottlenecked.[10] The value of $L_T$ can be estimated from plots of contact resistance ($2R_C$) verse $L_C$ at the point where the exponential increase begins.[10] As stated previously, $R_{CH}$ is negligible since our devices are fully contact dominated ($L_{ch}$ = 50 nm), therefore, $R_T \approx 2R_C$.[41] A plot of $2R_C$ verse $L_c$ at $n_s$ = 2.06×10$^{13}$ cm$^{-2}$ (**Fig. 5d**) demonstrates that back gate control (i.e., with CG) dramatically reduces the $L_T$ to below 30 nm, compared to $L_T \approx$ 100 nm without CG. This apparent improvement highlights how contact gating can artificially exaggerate the perceived scalability of $L_c$ when not accounted for in the analysis.[10,42,43]

**Conclusion**

In summary, a symmetric dual-gate monolayer MoS$_2$ FET structure with independently addressable top and back gates was used to perform a detailed study of the contact gating effect. This structure enables the direct extraction of a contact gating factor ($\beta_{CG}$) in both off- and on-state regimes, revealing that CG can enhance $I_{ON}$ and $g_m$ by up to ~2× in long-channel devices and by more than 5× in scaled architectures. Our analysis further shows that as channel and contact lengths are scaled, the device becomes increasingly contact-dominated, resulting in the amplification of CG on apparent performance metrics such as $R_C$, $L_T$, and $I_{ON}$.

These results underscore the critical role of device geometry in determining the performance potential of 2D FETs by demonstrating that CG can inflate key benchmarking parameters and obscure their true scaling limits. It's imperative to note that a device with full CG (i.e., gate overlapping source/drain electrodes) is not of relevance for a high-performance transistor technology as the overlap introduces parasitic capacitances between the gate and S/D contacts that unacceptably increase power consumption and intrinsic gate delays. What makes the observations in this work so important is the fact that the vast majority of reports on 2D FETs with ultralow contact resistance have been in FET geometries with full contact gating. While the symmetric dual-gate devices studied herein had relatively high contact resistance, they nevertheless provided evidence that contact gating has up to a 7× boost in on-state performance for scaled 2D FET geometries while also yielding a smaller transfer length – both of these improvements from CG should be acknowledged when benchmarking a reported 2D FET with record-setting performance metrics. Overall, eliminating CG effects is essential for accurately benchmarking the performance of 2D FETs and ensuring their viability for integration into high-performance transistor technologies.

## Methods

**Monolayer MoS$_2$ Film Growth**

The MoS$_2$ used throughout this study was grown via MOCVD on a 2'' double-side polished C-plane sapphire substrate at 1000°C. Molybdenum hexacarbonyl (Mo(CO)$_6$) and hydrogen sulfide (H$_2$S) were used as the metal and chalcogen precursors, respectively, while H$_2$ was used as the carrier gas. This sample was provided to us via the 2D Crystal Consortium (2DCC) as stated in the acknowledgements.

**Monolayer MoS$_2$ Film Transfer**

After local back gate patterning and deposition, followed by bottom dielectric and bottom seed layer deposition, the monolayer MoS$_2$ was transferred onto the target substrate using a PMMA-assisted wet transfer process.[44] The as-grown MoS$_2$ on sapphire was fixed to a glass slide with Kapton tape and spin-coated twice with 495k PMMA A6 at 4000 rpm for 45 seconds each. The coated sample was left for 24 hours to ensure good PMMA/MoS$_2$ adhesion. The film edges were then scratched with a razor blade, and the sample was removed from the glass slide before being immersed in 50 °C deionized (DI) water for 4 hours. During this immersion, water bubbles formed along the film edges, initiating delamination. The sample was subsequently removed from the bath and slowly reintroduced at an angle to promote complete delamination through capillary action. Due to the hydrophilic nature of the sapphire and the hydrophobic nature of MoS$_2$ and PMMA, capillary forces can effectively separate the PMMA/MoS$_2$ film from the substrate. After delamination, the floating film was retrieved using the target substrate. The substrate was then baked at 50 °C and 70 °C for 30 minutes each to remove residual water and improve adhesion. Finally, the PMMA support layer was dissolved in room temperature acetone for ~12 hours, followed by a 15-minute soak in isopropyl alcohol (IPA), yielding a clean monolayer MoS$_2$ film on the target substrate.

**Fabrication of Symmetric Dual-Gate FETs**

Device fabrication began on intrinsic Si substrates with a 1 μm thermally grown SiO$_2$ layer. Unless otherwise noted, all lithography steps followed the same procedure: substrates were spin-coated with 950k PMMA A5 resist (5 s at 500 rpm, 60 s at 3000 rpm), baked at 180 °C for 90 s, exposed by electron-beam lithography (EBL), and developed for 30 s in a 3:1 mixture of IPA and methyl isobutyl ketone (MIBK). Metal films were deposited by electron-beam evaporation, and all liftoff processes were carried out in 80 °C acetone for 2 h, followed by rinsing in IPA.

Ti/Au (10 nm/30 nm) alignment marks were first patterned by EBL and deposited, followed by liftoff. After alignment mark definition, local back gate (LBG) trenches were formed by EBL patterning/development, followed by RIE using CF$_4$/O$_2$ producing ~40 nm recesses in the SiO$_2$. Ni/Au (10 nm/30 nm) was then deposited to form the LBG electrodes, followed by liftoff. Larger Ti/Pd/Au (30 nm/30 nm/40 nm) pads were subsequently patterned, deposited, and lifted off for LBG probing contacts.

The bottom dielectric was deposited next via ALD of 25 nm HfO$_2$ (260 cycles) at 200 °C using tetrakis(dimethylamino)hafnium and H$_2$O as the metal precursor and oxidant, respectively.

A 1 nm Al seed layer was then deposited and naturally oxidized in air for 24 h to form an ultrathin $Al_2O_3$ layer, providing structural symmetry beneath the $MoS_2$ channel.

The monolayer $MoS_2$ was transferred onto the prepared substrate using a PMMA-assisted water transfer process as discussed above. Following transfer, S/D top contact leads for the $L_c$ and $L_{ch}$ scaling structures were spin-coated with 950k PMMA A2 resist (5 s at 500 rpm, 60 s at 3000 rpm) and patterned by EBL, followed by Ni/Au (5/5 nm) deposition and liftoff. Ti/Pd/Au (30 nm/30 nm/40 nm) pads were then patterned and deposited for S/D probing pads.

To define the $MoS_2$ channel width and ensure there was no electrical connection between the LBG, S/D, and top gate pads, an etching layer was patterned using EBL, and a brief $SF_6$ RIE (30 sccm, 100 W, 200 mTorr, 7 s). The sample was then placed in acetone to remove the resist. A top Al (1 nm) seed layer was subsequently deposited and oxidized under ambient conditions to promote nucleation on the $MoS_2$ for ALD. The top gate dielectric (25 nm $HfO_2$) was deposited under the same ALD conditions as the bottom dielectric. Finally, Ni/Au (10 nm/30 nm) top gate electrodes and Ti/Pd/Au (30 nm/30 nm/40 nm) top gate probing pads were patterned and deposited, completing the symmetric dual-gate $MoS_2$ FET structure. For a fabrication flow diagram, refer to **Supplementary Figure S2**.

**Scanning and Transmission Electron Microscopy**

Scanning electron microscopy (SEM) of the symmetrical double-gate monolayer $MoS_2$ FETs used in this study was conducted using an Apreo S by ThermoFisher Scientific at an accelerating voltage of 2.0 kV and a current of 25 pA (**Figure 1b-c**).

The sample for cross-sectional transmission electron microscopy (TEM) was prepared using a Scios Dual Beam Focused Ion Beam. High-angle annular dark-field scanning transmission electron microscopy (HAADF-STEM) and scanning TEM energy dispersive spectroscopy (EDS) were performed using a ThermoFisher Talos F200X with a Super-X EDS system and X-field emission gun source at an accelerating voltage of 200 kV (**Figure 1d-k**). For HAADF imaging, the detector's inner and outer collection semi-angles were set to 41 and 200 mrad, respectively, with a convergence semi-angle of 10.5 mrad.

**Electrical Measurements**

All electrical measurements were performed at room temperature in ambient conditions using the Lake Shore CRX-6.5K probe station connected to a Keysight B-1500 semiconductor parameter analyzer.

## Acknowledgements

We gratefully acknowledge support from the National Science Foundation under grants 2227175, 2401367, and 2328712. This work was primarily performed at the Duke University Shared Materials Instrumentation Facility (SMIF) (RRID:SCR_027480), a member of the North Carolina Research Triangle Nanotechnology Network (RTNN), which is supported by the NSF under award ECCS-2025064 as part of the National Nanotechnology Coordinated Infrastructure (NNCI). The authors gratefully acknowledge Roberto Garcia for STEM sample preparation and Chris Winkler for acquiring the STEM images at the Analytical Instrumentation Facility (AIF) at North Carolina State University. Materials support was provided by the NSF through the Pennsylvania State University 2D Crystal Consortium–Materials Innovation Platform (2DCC-MIP) under cooperative agreement DMR-2039351. The authors would also like to thank Rahul Banerjee for his assistance with obtaining the photoluminescence spectrum of the monolayer material used in this study.

## Author Contributions

V.M.R. and A.D.F. conceived the study. V.M.R. fabricated and tested the devices and analyzed the data. S.R.E contributed to the design of the figures. S.K.H assisted with electrical characterization of the devices. J.L.D. contributed to the experimental design. M.S.R. and T.R. transferred the monolayer MoS$_2$. A.D.F. supervised the research and provided scientific guidance. V.M.R. and A.D.F wrote the manuscript with revision and approval of all authors.

# Supplementary Information

**Impact of Contact Gating on Scaling of Monolayer 2D Transistors Using a Symmetric Dual-Gate Structure**

*Victoria M. Ravel[1], Sarah R. Evans[1], Samantha K. Holmes[1], James L. Doherty[1], Md. Sazzadur Rahman[1], Tania Roy[1], and Aaron D. Franklin[1,2]\**

[1]Department of Electrical & Computer Engineering, Duke University, Durham, NC 27708, USA

[2]Department of Chemistry, Duke University, Durham, NC 27708, USA

\*Address correspondence to aaron.franklin@duke.edu

## Supplementary Note 1: Materials Characterization of Monolayer $MoS_2$

To verify the quality and monolayer nature of the $MoS_2$ film, material characterization using Raman and photoluminescence (PL) spectroscopy was conducted. **Figure S1a-b**, respectively, show the Raman and PL spectra taken from the $MoS_2$ film. In **Figure S1a**, two prominent Raman peaks corresponding to the $E_{2g}^1$ (in-plane) and $A_{1g}$ (out-of-plane) vibrational modes are observed at ~386.58 cm$^{-1}$ and ~405.29 cm$^{-1}$, respectively. The peak-to-peak distance from these modes is ~18.71 cm$^{-1}$, representative of monolayer $MoS_2$.[1] The PL spectrum in **Figure S1b** shows an emission peak at ~1.82 eV, further confirming the monolayer nature of the film.[2]

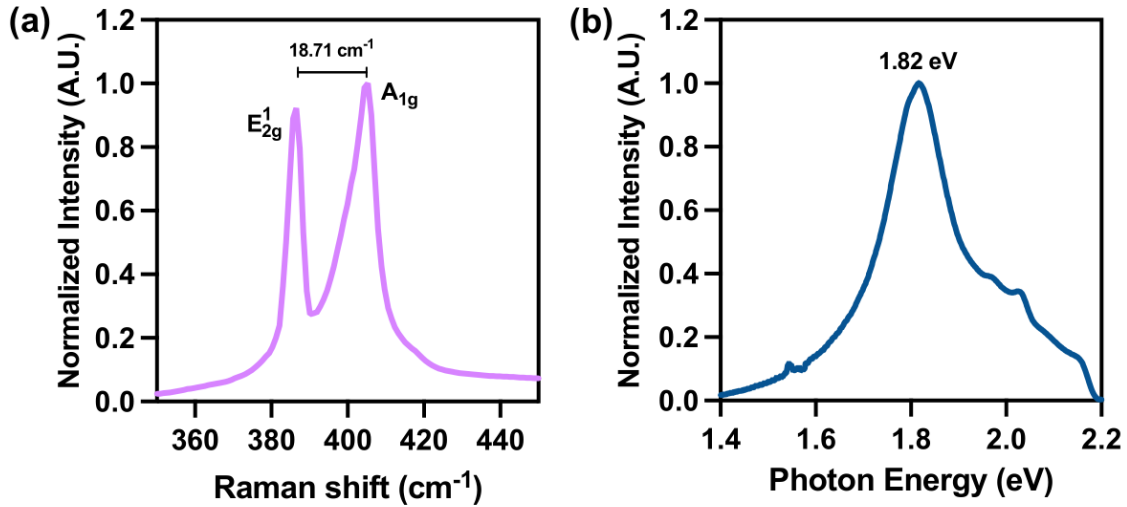

**Figure S1: Materials characterization of monolayer $MoS_2$.** **(a)** Raman and **(b)** photoluminescence (PL) spectra confirming the monolayer nature of the MOCVD-grown $MoS_2$ film used in this study. Raman and PL spectra were collected using 442 nm and 532 nm excitation lasers, respectively.

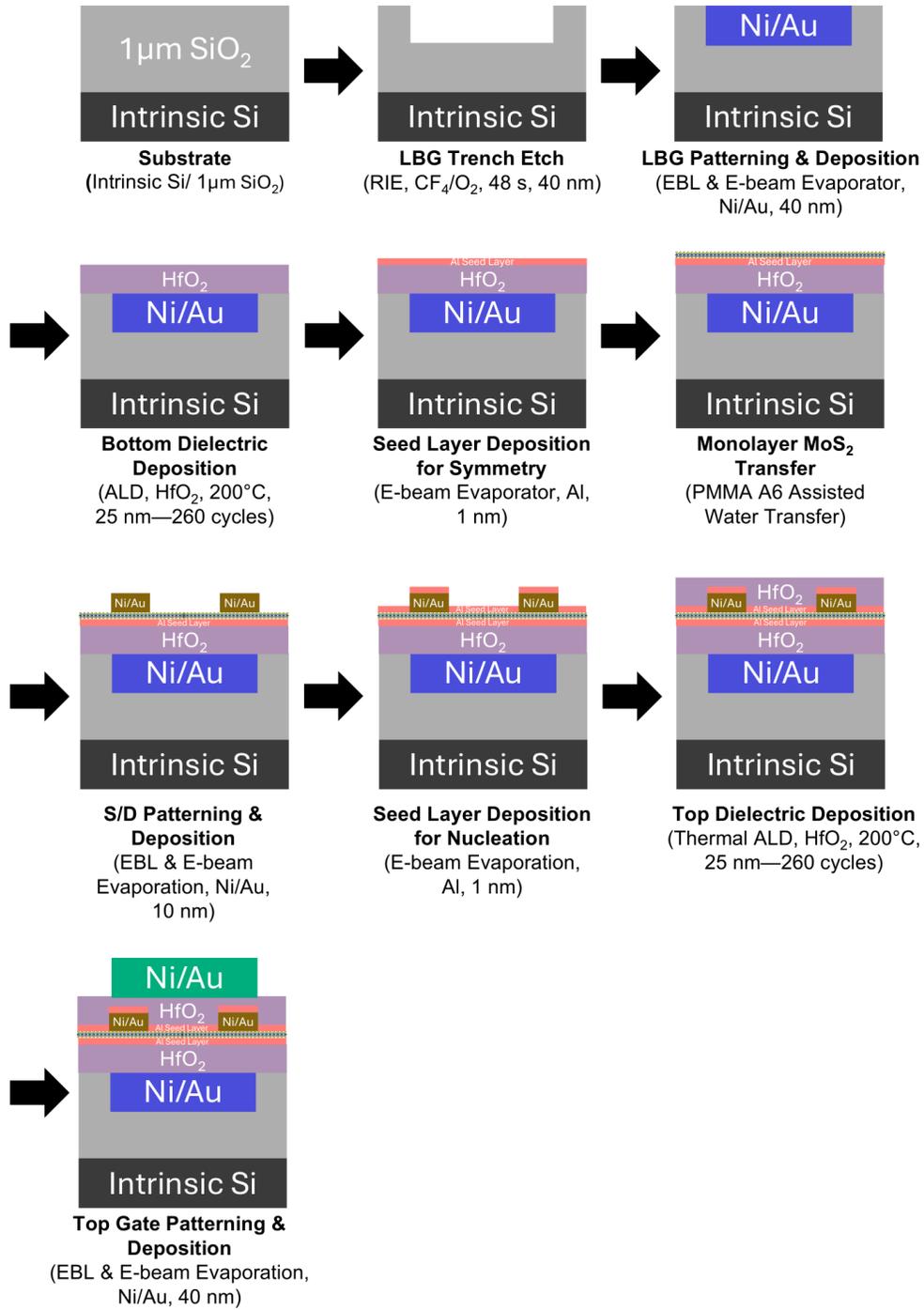

**Figure S2: Symmetric double-gated transistor fabrication flow.** Figures showing the cross-section of one device in the channel region.

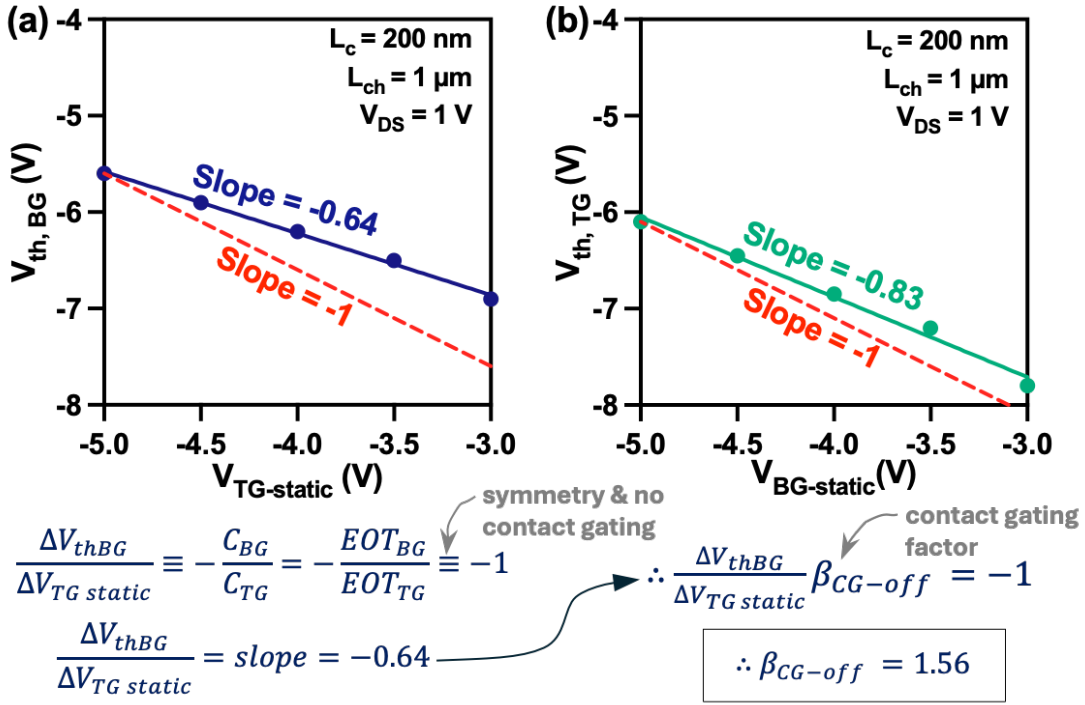

**Figure S3: Impact of contact gating on off-state threshold voltage for the device shown in Fig. 2, quantified by $\beta_{CG\text{-off}}$.** (a) Back gate threshold voltage, $V_{th,\,BG}$, extracted at iso-current of $I_D = 10$ nA/μm, as a function of static gate $V_{TG}$. Since $V_{BG}$ is swept, contact gating is fully present in the $V_{th,\,BG}$ extraction. (b) Same as (a), but the top gate threshold voltage, $V_{th,\,TG}$, is plotted vs. $V_{BG}$, where only a fixed (static) contact gating influence is present, yielding a slope closer to -1. Calculation of the off-state contact gating factor is shown.

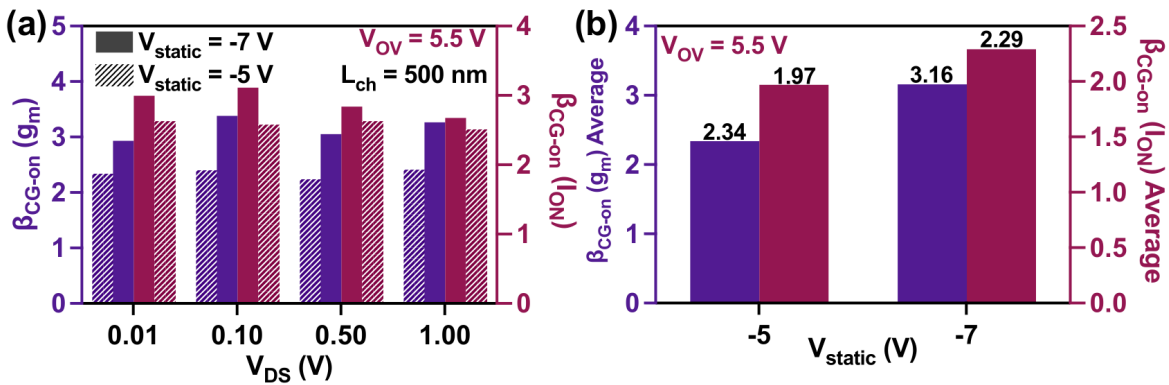

**Figure S4. $I_{on}$ and $g_m$ boost from contact gating under various bias conditions.** (a) $\beta_{CG\text{-on}}$ ($I_{ON}$) and $\beta_{CG\text{-on}}$ ($g_m$) vs. $V_{DS}$ for $V_{static} = -5$ and $-7$. (b) Average $\beta_{CG\text{-on}}$ ($I_{ON}$) and $\beta_{CG\text{-on}}$ ($g_m$) for all $V_{DS}$ in (a).

**Supplementary Note 2: Contact resistance comparison between back and top gate devices at $V_{static}$ = -5 V and -7 V**

For the channel length ($L_{ch}$) scaling structure (TLM structure) shown in **Figure 1c** of the manuscript, transfer characteristics were measured under both top- and back gate operation at various $V_{static}$ values. The total resistance ($R_T = \frac{V_{DS}}{I_D}$) was plotted as a function of $L_{ch}$ for different gate overdrive voltages ($V_{OV} = V_{GS} - V_{th}$) (**Figure S5 and S6**). **Figures S5a and S6a** show $R_T$ versus $L_{ch}$ for devices operated with back gate control at $V_{static}$ = -5 V and -7 V, respectively. **Figures S5b** and **S6b** present the corresponding data with the top gate modulating the channel region. The contact resistance ($2R_C$) was obtained from the y-intercept of these plots and subsequently plotted as a function of sheet carrier density ($n_s = \frac{C_{ox}V_{ov}}{q}$) in **Figures S5c** and **S6c**. The gate oxide capacitance ($C_{OX} = \frac{\varepsilon_{ox}\varepsilon_0}{t_{ox}}$) was calculated by modeling the hafnia and alumina layers as capacitors connected in series ($\frac{1}{C_{OX}} = \frac{1}{C_{HfO_2}} + \frac{1}{C_{Al_2O_3}}$), where $\varepsilon_{Al_2O_3} \approx 7$ and $\varepsilon_{HfO_2} \approx 15$, yielding $C_{OX}$ = 4.89×10$^{-7}$ F·cm$^{-2}$. At a carrier concentration of $n_s$ = 2.06×10$^{13}$ cm$^{-2}$, it is found that increasing the magnitude of the negative $V_{static}$ value amplifies the difference between the $2R_C$ values extracted top- and back gate operation. At $V_{static}$ = -5 V, the top- and back gate $2R_C$'s differ by ~61%, while at $V_{static}$ = -7 V this difference increases to ~87% difference. The back gate $2R_C$ value remains nearly consistent between these static biases, whereas the top gate contact resistance more than doubles as $V_{static}$ shifts from -5 V to -7 V. This increase occurs because, at $V_{static}$ = -7 V, the bias lies deep in the subthreshold regime, leading to severe depletion of the MoS$_2$ beneath the contacts when the back gate is held static, which results in an order of magnitude increase in top gate contact resistance.

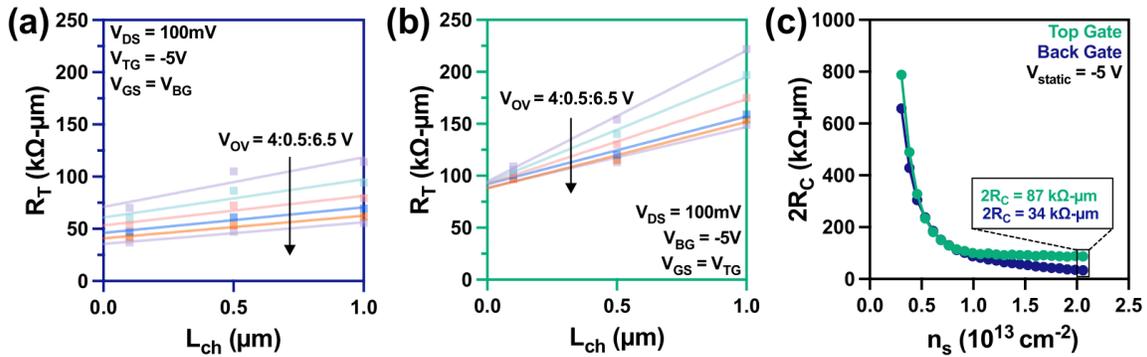

**Figure S5. Contact resistance comparison of top and back gate control at $V_{static}$ = -5 V. (a,b)** Transfer length method (TLM) plots showing the total resistance ($R_T = R_{CH} + 2R_C$) for **(a)** back gate and **(b)** top gate devices at a static gate voltage of -5 V, and across different overdrive voltages ($V_{OV}$) on the control gate. **(c)** $2R_C$ vs. the sheet carrier density of the top and back gate devices, showing that at $V_{static}$ = -5 V, there is an improvement in $R_C$ with full contact gating present (i.e., back-gated, blue curve).

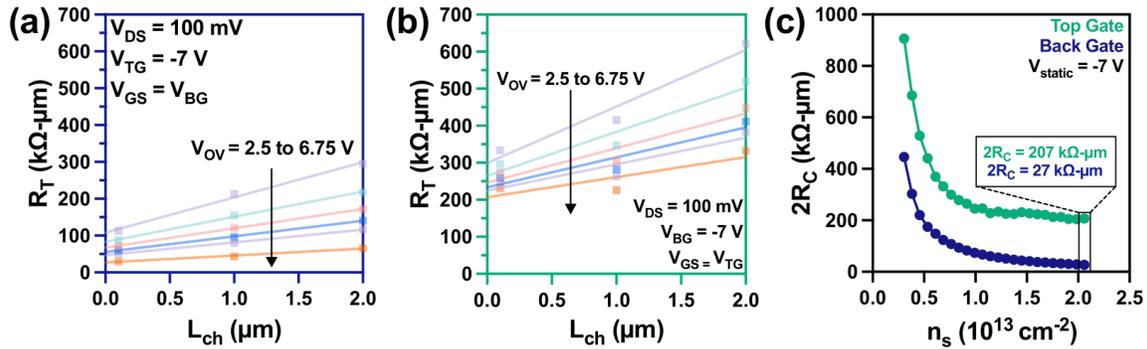

**Figure S6. Contact resistance comparison of top and back gate control at $V_{static}$ = -7 V.** TLM plot of **(a)** back gate and **(b)** top gate devices at a static gate voltage of -7 V (off-state), and across different $V_{OV}$ on the control gate. **(c)** $2R_C$ vs. the sheet carrier density of the top- and back gate devices, showing that at $V_{static}$ = -7 V, there is a dramatic increase in $R_C$ for the top-gated device as the contact gating from the $V_{static}$ depletes the contacted $MoS_2$ regions.

**Supplementary Note 3: Contact gating effects on field-effect mobility**

The field effect mobility was extracted using

$$\mu_{FE} = \frac{g_m L_{ch}}{W C_{ox} V_{DS}}$$

where W is the channel width, $g_m$ is the transconductance, $L_{ch}$ is the channel length, $C_{ox}$ is the oxide capacitance and $V_{DS}$ is the drain to source voltage.

The transconductance was determined by averaging the slope over 11 points by taking 5 points above and below the steepest slope point on each transfer curve. For the 2 μm channel length device, μ$_{FE}$ decreases by 77%, from $10.7 \frac{cm^2}{V \cdot s}$ to $2.47 \frac{cm^2}{V \cdot s}$, when the top gate is swept while the back gate's V$_{static}$ value is reduced from 0 V to -7 V. In contrast, sweeping the back gate yields only a 4.1% reduction, from $12.2 \frac{cm^2}{V \cdot s}$ to $11.6 \frac{cm^2}{V \cdot s}$, over the same static bias range applied on the top gate. As the L$_{ch}$ is scaled down to 100 nm, μ$_{FE}$ loses physical significance because carrier transport becomes dominated by scattering, with the channel length approaching the mean free path.[3]

The observed μ$_{FE}$ trends align with the contact resistance trends shown in Figures S5 and S6: as the V$_{static}$ applied on the back gate becomes more negative (deeper into the subthreshold region of the devices), R$_C$ increases by orders of magnitude, which suppresses g$_m$ and consequently reduces the extracted μ$_{FE}$. This suppression of g$_m$ is evident when comparing Figure 2c and 2e in the main manuscript.

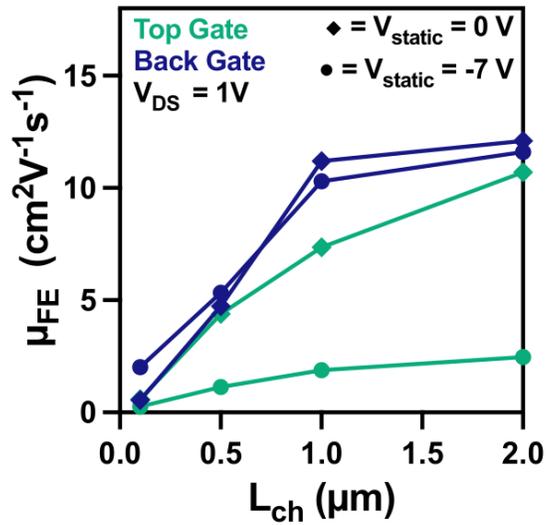

**Figure S7.** Field-effect mobility versus channel length at $V_{static}$ = 0 V, -7 V, showing the impact of CG-depleted MoS$_2$ in contact regions on $\mu_{FE}$.

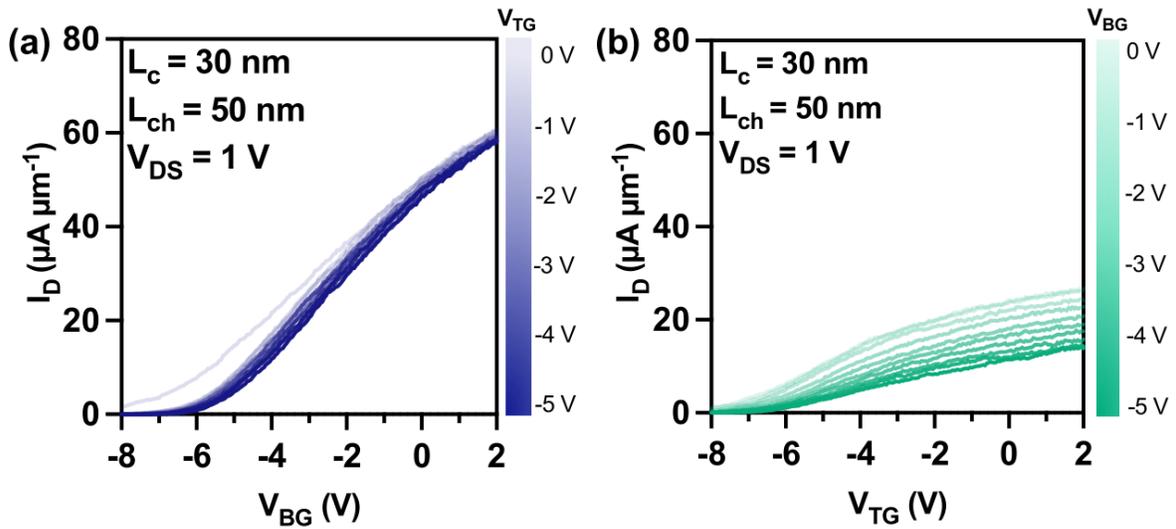

**Figure S8. Transfer characteristics of the device analyzed in Fig. 5(c).** (a) Back gate swept with a static top gate bias ranging from -5 V to 0 V, and (b) top gate swept with a static back gate bias from -5 V to 0 V.

**Supplemental Information References:**